\documentclass[12pt]{iopart}

\usepackage{bm}
\usepackage{epsfig}
\usepackage{citesort}
\usepackage{graphicx}
\eqnobysec

\newcommand{\lambdabar}%
{{\hbox{$\lambda$\kern-1.ex\raise+0.45ex\hbox{--}}}}

\long\def\dump#1{}

\begin{document}

%%%%%%%%%%%%%%%%%%%%%%%%%%%%%%%%%%%%%%%%%%%%%%%%%%%%%%%%%%%%%%%%%%%%%%
% Front page %%%%%%%%%%%%%%%%%%%%%%%%%%%%%%%%%%%%%%%%%%%%%%%%%%%%%%%%%%
%%%%%%%%%%%%%%%%%%%%%%%%%%%%%%%%%%%%%%%%%%%%%%%%%%%%%%%%%%%%%%%%%%%%%%

\begin{flushright}
{\large \tt MPP-2007-134 \\
TUM-HEP-674/07}
\end{flushright}

\title{Effects of CMB temperature uncertainties on cosmological
parameter estimation}

\author{Jan~Hamann$^1$ and Yvonne~Y.~Y.~Wong$^2$}

\address{$^1$~Physik Department T30e,
 Technische Universit\"at M\"unchen\\
 James-Franck-Strasse, D-85748 Garching, Germany\\
 $^2$~Max-Planck-Institut f\"ur Physik (Werner-Heisenberg-Institut)\\
 F\"ohringer Ring 6, D-80805 M\"unchen, Germany}

\ead{\mailto{jan.hamann@ph.tum.de},
     \mailto{ywong@mppmu.mpg.de}}

\begin{abstract}
We estimate the effect of the experimental uncertainty in the
measurement of the temperature of the cosmic microwave background
(CMB) on the extraction of cosmological parameters from future CMB
surveys. We find that even for an ideal experiment limited only by
cosmic variance up to $\ell = 2500$ for both the temperature and
polarisation measurements, the projected cosmological parameter errors
are remarkably robust against the uncertainty of 1 mK in the {\sc
firas} CMB temperature monopole measurement.  The maximum degradation
in sensitivity is $20\%$, for the baryon density estimate, relative to
the case in which the monopole is known infinitely well. While this
degradation is acceptable, we note that reducing the uncertainty in
the current temperature measurement by a factor of five will bring it
down to the per cent level.  We also estimate the effect of the
uncertainty in the dipole temperature measurement. Assuming the
overall calibration of the data to be dominated by the dipole error of
$0.2\%$ from {\sc firas}, the sensitivity degradation is insignificant
and does not exceed 10\% in any parameter direction.

\end{abstract}
\maketitle

%%%%%%%%%%%%%%%%%%%%%%%%%%%%%%%%%%%%%%%%%%%%%%%%%%%%%%%%%%%%%%%%%%%%%%
\section{Introduction}\label{sec:introduction}

%%%%%%%%%%%%%%%%%%%%%%%%%%%%%%%%%%%%%%%%%%%%%%%%%%%%%%%%%%%%%%%%%%%%%%

The precision measurement of the energy spectrum of cosmic microwave
background (CMB) photons by the {\sc firas} instrument \cite{FIRAS} on
board the {\sc {\sc cobe}} satellite \cite{COBE} is one of the most
spectacular pieces of evidence in support of the big bang
theory. Owing to the fact that cosmological expansion does not lead to
spectral distortions, but merely shifts the spectrum towards longer
wavelengths, today's CMB spectrum still resembles that of a black body
of temperature $T_0$ \cite{Fixsen:1996nj}, even though the photons
have not been in thermal equilibrium since last scattering. 
As it happens, $T_0$ is one of the few cosmological parameters that
are accessible to direct measurement, without the need to resort to 
the model-dependent process of statistical inference.  Combining the
results of three independent estimation methods, the authors of
reference~\cite{FixMat} (see also \cite{Mather:1998gm}) find $T_0 =
2.725 \pm 0.001 \ {\rm K}$ (at 68\% c.l.), an impressive accuracy
$\Delta T_0/T_0$ of better than $0.04\%$.

Since the CMB monopole $T_0$ determines the present radiation density
of the Universe, it is also a fundamental input parameter for the
calculation of the temperature and polarisation anisotropies of the
CMB. Naturally, the experimental error in $T_0$ will also introduce a
theoretical uncertainty $\Delta \mathcal{C}_\ell/\mathcal{C}_\ell$ in
the prediction of the angular power spectra
$\mathcal{C}_\ell$. Depending on the scale, this uncertainty can reach
a magnitude of order a few times $0.1\%$ \cite{Hu:1995fq,Chluba:2007zz}.

If one wants to use the observed $\mathcal{C}_\ell$ data to infer
constraints on the free parameters of a particular cosmological model,
this temperature effect ought, in principle, to be taken into
account. In a statistically stringent Bayesian analysis, one would
have to treat $T_0$ as a free parameter and impose a suitable prior on
its value instead of keeping it fixed.
Given the statistical errors of present CMB anisotropy data
\cite{Hinshaw:2006ia,Page:2006hz}, current parameter estimates are
unlikely to be affected. However, in the near future, experiments such
as {\sc planck} \cite{planck} or {\sc cmbpol} \cite{Bock:2006yf} will
be able to measure the temperature and polarisation angular power
spectra to an accuracy that is essentially limited by cosmic variance
over a wide range of multipoles up to $\ell \sim 2500$. It is
therefore timely to ask whether parameter estimates using these high
quality data sets may be compromised by a possibly insufficiently
accurate measurement of $T_0$.

In addition to the monopole, the analysis and interpretation of the
data taken by full-sky CMB experiments also depends on the {\sc firas}
measurement of the temperature amplitude of the CMB dipole, $\tau_{\rm
dp} = 3.381 \pm 0.007 \ {\rm mK}$ \cite{FixMat}. Unlike $T_0$, the
dipole temperature is not important for the theoretical prediction of
the power spectra, but affects the experimental values of the
$\mathcal{C}_\ell\,$s. The dipole provides a convenient way to
calibrate detector output with the amplitude scale of fluctuations
in temperature and polarisation. For {\sc wmap}
\cite{{Hinshaw:2003fc}} and the low- and high-frequency instruments of
{\sc planck} \cite{Bersanelli,Piat:2001he}, the error in the absolute
calibration will be limited by the uncertainty in $\tau_{\rm dp}$,
inducing a normalisation uncertainty in the angular power spectra data
of $\Delta \mathcal{C}^{\rm exp}_\ell/\mathcal{C}^{\rm exp}_\ell = 2
\Delta \tau_{\rm dp}/\tau_{\rm dp} \simeq 0.4\%$
\cite{Cappellini:2003bh}.   

In the present work, we determine how large an effect the
uncertainty in the values of $T_0$ and $\tau_{\rm dp}$ will have on
the estimates of cosmological parameters for the analysis of future
CMB data. In section \ref{sec2} we outline the r\^ole played by $T_0$
in the calculation of the anisotropies of the CMB. In section
\ref{sec3} we describe the technical details of our analysis, the
results of which are presented in section \ref{sec4}. We summarise our
results and conclude in section \ref{sec5}. A detailed account of the
technicalities of generating mock CMB data is given in the
Appendix.

%%%%%%%%%%%%%%%%%%%%%%%%%%%%%%%%%%%%%%%%%%%%%%%%%%%%%%%%%%%%%%%%%%%%%%
\section{CMB temperature and the anisotropy spectra \label{sec2}}
%%%%%%%%%%%%%%%%%%%%%%%%%%%%%%%%%%%%%%%%%%%%%%%%%%%%%%%%%%%%%%%%%%%%%%

The present temperature of the cosmic microwave background is one of
the basic input parameters for the calculation of anisotropies,
affecting the evolution of the fluctuations during various stages of
the early Universe.
In particular, it determines directly the current photon energy density
via 
\begin{equation}
	\rho_{\gamma,0} = \frac{\pi^2}{15} \, T_0^4.
\end{equation}
Let us sketch briefly how the calculation of the angular power spectra
will explicitly depend on $T_0$ or $\rho_{\gamma,0}$, and to what
extent these effects can mimic changes in other free parameters of the
cosmological model.

\paragraph{Baryon-to-photon ratio}

A fundamental input parameter in the Boltzmann equations for the
baryon density perturbations is the baryon-to-photon density ratio,
\begin{equation}
R \equiv \frac{3 \rho_{\rm b}}{4 \rho_\gamma}.
\end{equation}
Thus, already at the level of the perturbations equations, there
exists an exact degeneracy between $T_0$ and the physical baryon
density  $\omega_{\rm b} \equiv \Omega_{\rm b} h^2$.
In the tight-coupling limit valid before recombination, $R$ defines
the  sound speed for the coupled baryon--photon fluid via
$c_\mathrm{s}^2 \equiv 1/3 \left(1+R \right)$, and enhances the
compression phase (hence alternate peaks) of the acoustic oscillations
\cite{Hu:1994uz}. The comoving sound horizon
\begin{equation}
\label{eq:soundhorizon}
r_s (\eta_*) \equiv \int^{\eta_*}_0 d \eta \ c_s(\eta)
\end{equation}
evaluated at the time of recombination $\eta_*$ governs the spacing of
the acoustic peaks in the observed CMB anisotropies.  The suppression
of the anisotropy spectra at high $\ell$ due to diffusion damping also
depends explicitly on $R$.

\paragraph{Recombination}

The details of the process of recombination
\cite{Peebles:1968ja,Zeldovich:1969en}, during which the photons
decouple from the plasma, evidently have a significant influence
on the eventual CMB anisotropies. Between redshifts $900 < z < 1500$, the
free electron fraction $X_\mathrm{e}$ can be approximated by
\cite{Sunyaev:1970eu}
\begin{equation}
	N_\mathrm{e} \propto \frac{h \, T_0^{1/2}}{z
\sqrt{\Omega_{\rm m}}} \,\exp \left[ - \frac{B}{z\,T_0} \right],
\end{equation}
where $h$ is the dimensionless Hubble parameter today,
$\Omega_\mathrm{m}$ is the matter density, and $B \simeq 3.9 \times
10^4 \mathrm{\ K}$ is a numerical constant.
Because of the exponential dependence on the temperature, we can expect
$\Delta N_\mathrm{e}/N_\mathrm{e} \gg \Delta T_0/T_0$. From a more
sophisticated calculation, it was shown in reference \cite{Chluba:2007zz}
that $\Delta N_\mathrm{e}/N_\mathrm{e}$ can be as large as $0.55\%$
for $\Delta T_0/T_0 \sim 0.04\%$
within the standard $\Lambda$CDM model.

\paragraph{Matter-radiation equality}

The parameter $T_0$ determines not only the photon energy density, but
also, implicitly, the neutrino energy density, and thus the total radiation
density before the neutrinos become non-relativistic:
\begin{equation}
	\rho_\mathrm{r} = \frac{\pi^2}{15} \, T_0^4 \, (1+z)^4 \left[ 1 +
\frac{7}{8} \, N_\mathrm{eff} \left( \frac{4}{11} \right)^{4/3}\right].
\end{equation}
Assuming the particle content of the standard model and standard
neutrino decoupling, $N_\mathrm{eff} \simeq 3.046$ \cite{Mangano:2005cc}.
It is apparent that any change in $\rho_\mathrm{r}$ will shift the
time of matter--radiation equality $z_\mathrm{eq}$, 
which is manifest in an enhancement especially of the first acoustic peak
relative to the low $\ell$ plateau
through the early integrated Sachs--Wolfe effect.  Thus,
one can expect some degree of degeneracy between $T_0$ and the
physical matter density $\omega_{\rm m} \equiv \Omega_{\rm m} h^2$ in
the CMB anisotropies.

\paragraph{Projection}

The projection of the temperature and polarisation fluctuations onto the 
sky introduces for the observed CMB anisotropies an additional dependence 
on the angular diameter distance $D_*$ to the last scattering surface.  For
a flat geometry, 
\begin{equation}
D_* = \int^1_{a_*} \frac{d a}{a^2 H(a)},
\end{equation}
where 
$H(a)=100 \sqrt{h^2 + \Omega_{\rm m} h^2 (a^{-3} -1) + \Omega_{\rm r} h^2 (a^{-4}-1)}$.
Evidently, projection leads to a degeneracy between $h$ and $T_0$, both directly, and 
indirectly through $\omega_{\rm m}$'s correlation with $T_0$.

\bigskip

\begin{figure}[t]
%\begin{center}
\includegraphics[height=.49\textwidth,angle=270]{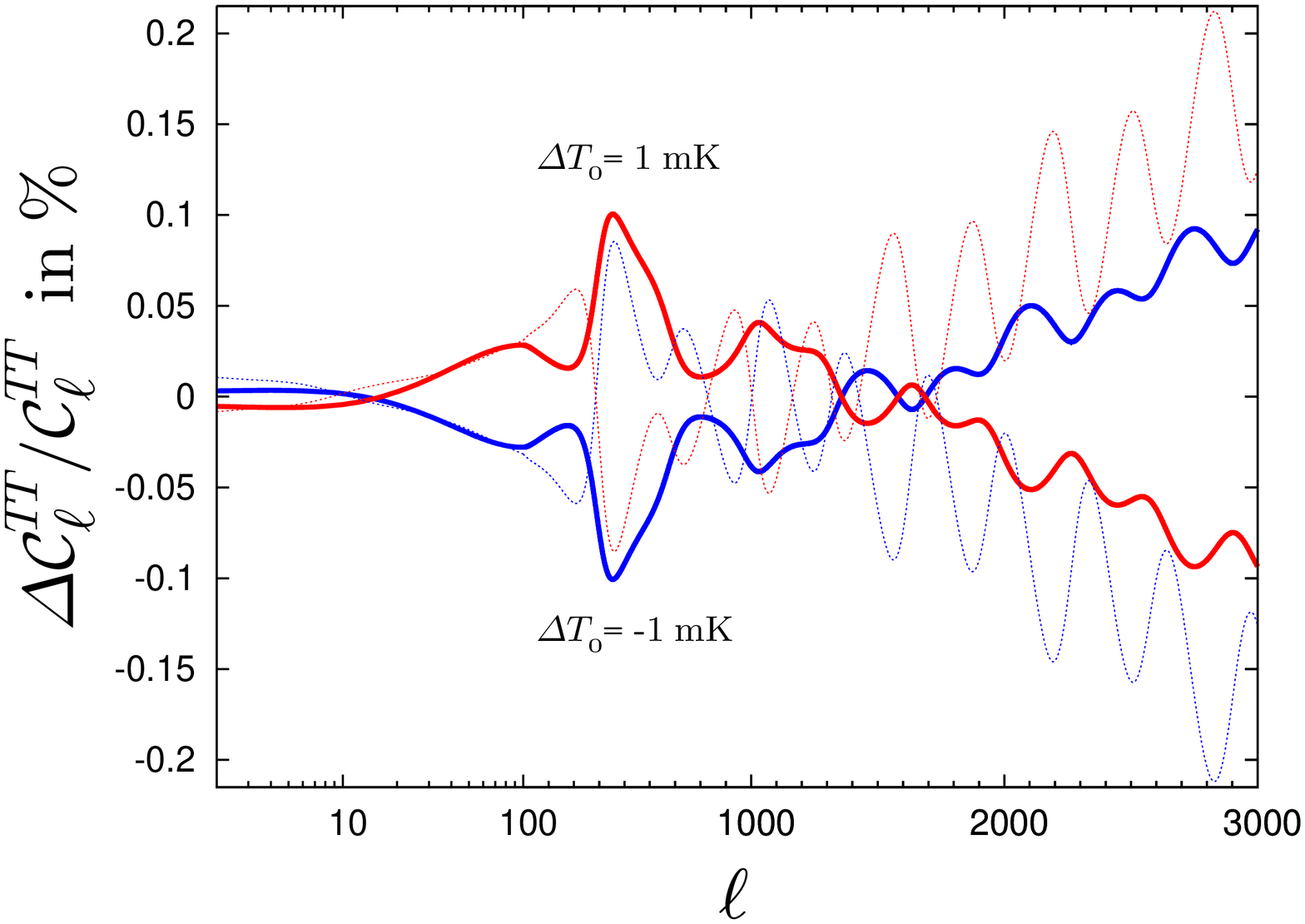}
\includegraphics[height=.49\textwidth,angle=270]{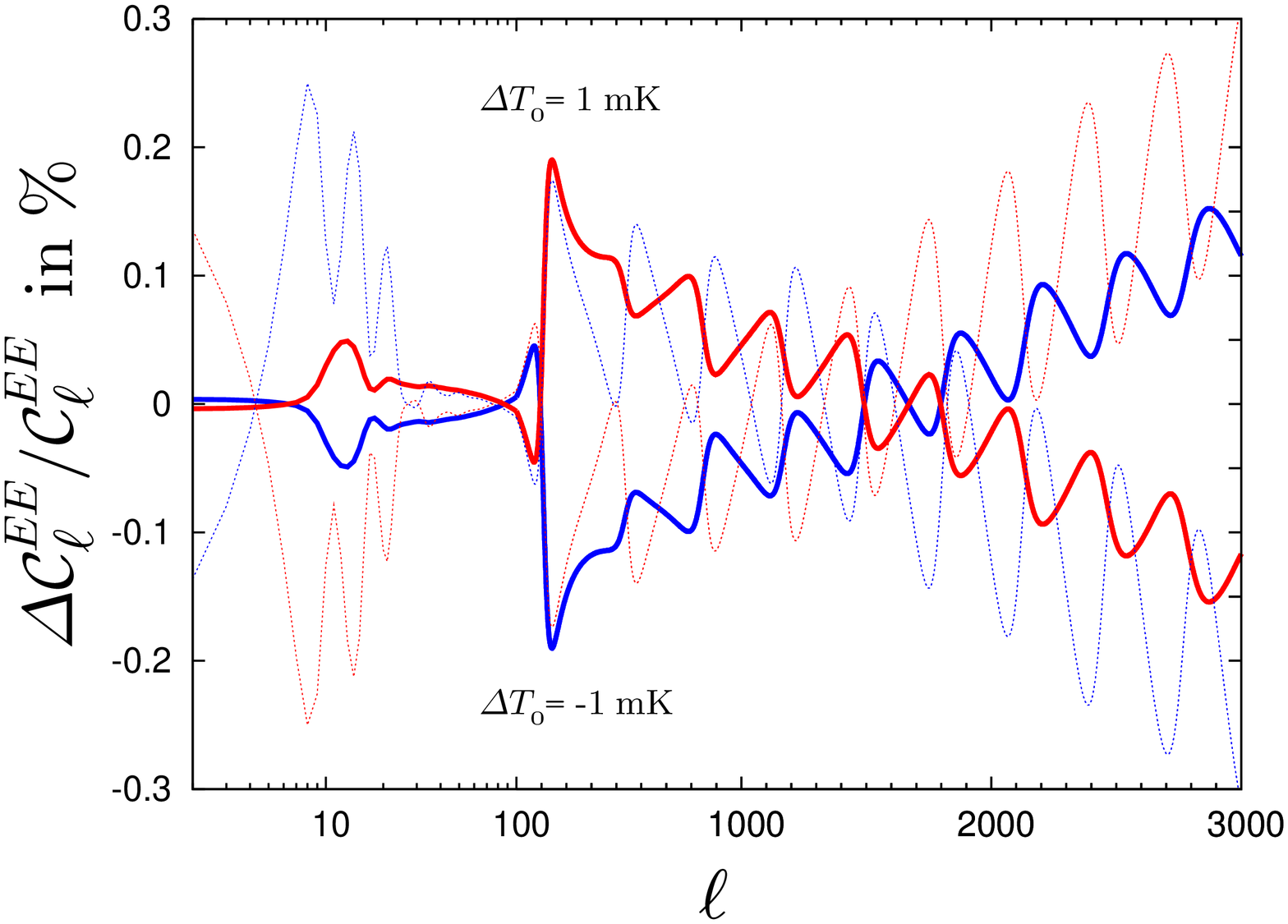}
\caption{\label{fig:cls}
These diagrams show the difference between the angular power spectra
under a change in $T_0$ of $\pm 1 {\rm mK}$ (thick solid lines). We
also plot the effects of changing the baryon density $\omega_{\rm b} =
\Omega_{\rm b} h^2$ by $\Delta \omega_{\rm b}/\omega_{\rm b} = 4 \,
\Delta T_0/T_0$, holding $T_0$ fixed (thin dotted lines). \emph{Left
panel}: Temperature autocorrelation. \emph{Right panel}:
Polarisation $E$-mode autocorrelation.} 
%\end{center}
\end{figure}

We conclude from this brief discussion that the parameter most degenerate 
with $T_0$ is the baryon density $\omega_{\rm b}$, followed 
by the matter density $\omega_m$ and the Hubble parameter $h$. 
Combining these three effects leads to an uncertainty in the angular power spectra 
of $\Delta \mathcal{C}_\ell/\mathcal{C}_\ell \sim 0.2\%$ for $\Delta T_0/T_0 \sim 0.04\%$.
For the fiducial model of section \ref{sec3} we
illustrate this uncertainty in figure \ref{fig:cls}.

%%%%%%%%%%%%%%%%%%%%%%%%%%%%%%%%%%%%%%%%%%%%%%%%%%%%%%%%%%%%%%%%%%%%%%
\section{Methodology\label{sec3}}
%%%%%%%%%%%%%%%%%%%%%%%%%%%%%%%%%%%%%%%%%%%%%%%%%%%%%%%%%%%%%%%%%%%%%%

In order to answer the question whether the standard approach of
keeping $T_0$ fixed will be justified when fitting future data, we
compare the results of an analysis that treats $T_0$ and $\tau_{\rm
dp}$ as free (albeit well-constrained) parameters with those of a fit
where these two parameters are kept constant.

\subsection{Mock data sets}
Following the method outlined in reference~\cite{Perotto:2006rj}, we
generate two sets of mock CMB anisotropy data, comprising the
$TT$, $TE$ and $EE$ angular power spectra for multipoles $2 \leq \ell
\leq 2500$, assuming respectively the projected noise levels of the
{\sc planck} experiment (see table~\ref{table:cmbexp}), and an ideal,
noiseless cosmic variance limited experiment ({\sc cvl}). We assume a
sky coverage of $f_{\rm sky} = 0.65$ in both cases. For a more
detailed discussion of the method, we refer the reader to
\ref{sec:appendix}. Our fiducial model is specified by the parameter
values listed in table~\ref{table:params}.

\begin{table}[t]
\caption{\label{table:cmbexp}
List of the experimental parameters assumed for the {\sc planck}
satellite \cite{planck}: $\theta_{\rm beam}$ measures the
width of the beam, $\Delta_{T,P}$ are the sensitivities per pixel and
$\nu$ is the centre frequency of the channels.}\vskip5mm \hskip45mm
\footnotesize{
\begin{tabular}{cccc}
\br
 $\nu$/GHz & $\theta_{\rm beam}$ & $\Delta_T$/$\mu$K & $\Delta_P$/$\mu$K\\
\mr
 100 & 9.5' & 6.8 & 10.9\\
 143 & 7.1' & 6.0 & 11.4\\
 217 & 5.0' & 13.1 & 26.7\\
\br
\end{tabular}}
\end{table}

In the generation as well in the subsequent analysis of the data we
use the recombination code \texttt{recfast}
\cite{Seager:1999km,Seager:1999bc} and ignore secondary effects such
as gravitational lensing or the Sunyaev-Zel'Dovich effect. Let us
stress that real data would require a less simplistic treatment of the
physics of recombination (see, e.g.,
references~\cite{Dubrovich:2005fc,Chluba:2006bc}) and the secondary
effects; failure to do so can severely bias results
\cite{Lewis:2006ym}.

\subsection{Parameter estimation}

The multi-dimensional posterior probability distributions
$\mathcal{P}(\theta)$ are reconstructed using a modified version of
\texttt{CosmoMC} \cite{Lewis:2002ah}, a Markov Chain Monte Carlo
(MCMC) algorithm used in conjunction the \texttt{CAMB}
\cite{Lewis:1999bs} code to calculate polarisation and temperature
spectra. For each analysis we generate eight Markov chains; their
convergence is monitored using the Gelman and Rubin $R$-parameter
\cite{gelru}. Our convergence criterion is $R-1 < 0.01$, a much
stricter requirement than for instance the one used by the {\sc wmap}
team
\cite{Spergel:2006hy}.  

We analyse two basic models: the widely-used six-parameter ``vanilla''
model, and an extended model (vanilla+$Y_{\rm He}$), where in addition
we vary the primordial helium fraction. With the exception of $T_0$
and $\tau_{\rm dp}$, we impose flat top-hat priors on the free
parameters of the models; the limits are listed in table
\ref{table:params}.

For each of the models we perform the analysis with the CMB
temperature and dipole either kept fixed at their fiducial values, or
treated as free parameters.
We account for the experimental error in $T_0$ by imposing a Gaussian
prior of the form
\begin{equation}
	\pi(T_0) \propto \exp\left[-\frac{1}{2} \left(\frac{T_0 - 2.725\
\mathrm{K}}{0.001\ \mathrm{K}}\right)^2\right].
\end{equation}
The dipole, in principle, has no effect on the theoretical prediction,
only on the data. However, since the absolute calibration affects 
polarisation and temperature data in the same way, on all scales, we
do not need to generate new data each time $\tau_{\rm dp}$
changes. Instead, we shift the calibration uncertainty to the theory
side, by substituting the normalisation of the primordial power
spectrum with
\begin{equation}
	\label{eq:diponorm}
	A_{\rm S} \longrightarrow A_{\rm S} \; \left(1 + 2 \;
\frac{\tau_{\rm dp} - 3.381\ \mathrm{mK}}{3.381\ \mathrm{mK}}\right),
\end{equation} 
with a Gaussian prior on $\tau_{\rm dp}$,
\begin{equation}
	\pi(\tau_{\rm dp}) \propto \exp\left[-\frac{1}{2} \left(\frac{\tau_{\rm dp} - 3.381\
\mathrm{mK}}{0.007\ \mathrm{mK}}\right)^2\right],
\end{equation}
corresponding to the result of the {\sc firas} measurement.

\begin{table}[t]
%\begin{center}
\caption{In this table we show the free parameters of our model,
their fiducial values used to generate the mock data and the prior
ranges adopted in the analysis.\label{table:params}}\vskip5mm \hskip15mm
\footnotesize{\begin{tabular}{llll}
 \br
 Parameter&&Fiducial Value&Prior Range\\
 \mr
 Dark matter density & $\Omega_{\rm dm} h^2$ & 0.104 & $0.01\to0.99$ \\
 Baryon density & $\Omega_{\rm b} h^2$ & 0.0223  & $0.005 \to 0.1$ \\
 Hubble parameter & $h$ & 0.7 & $0.4\to1$ \\
 Redshift of reionisation & $z_\mathrm{re}$  & 11.26 & $3\to25$\\
 Normalisation @ $k=0.002\mathrm{\ Mpc}^{-1}$&
$\ln(10^{10}A_\mathrm{S})$ & 3.135 & $2.7\to4$\\ 
 Scalar spectral index & $n_\mathrm{S}$ & 0.96  & $0.5\to2$\\
 Helium fraction & $Y_\mathrm{He}$ & 0.24 & $0.1\to0.4$\\
 CMB temperature & $T_0$ & 2.725 K & see text\\
 CMB dipole & $\tau_{\rm dp}$ & 3.381 mK & see text\\
 \br
\end{tabular}}
%\end{center}
\end{table}

Note that we avoid using the popular Kosowsky-parameter $\theta_s$
\cite{Kosowsky:2002zt}, defined as the ratio of the sound horizon 
at recombination to the angular diameter distance to the last 
scattering surface, and fit instead the Hubble parameter $h$ directly. 
Mapping between $\theta_s$ and $h$ as implemented in \texttt{CosmoMC}
involves the use of a fitting formula to determine the recombination
redshift $z_*$, which was derived in reference \cite{Hu:1995en} under
a number of assumptions, including that of a fixed temperature, and is
thus not applicable in our analysis.

Since the future data sets considered here will be able to constrain
the parameters of these models extremely well, one can expect the
resulting posterior distribution to be reasonably close to a
multivariate Gaussian near its mode. As a consequence, adding an
additional parameter, such as $T_0$, with a Gaussian posterior, is
unlikely to shift the point estimates of other parameters. Also, one
would not expect the errors of uncorrelated parameters to be affected;
only parameters that are degenerate with the new parameter are likely
to have increased error bars. Naturally, this assumes that the
Gaussian is actually centred around the fixed value -- using a wrong
value of the temperature will of course bias the best fit.

The expected near-Gaussianity of the posterior also implies that
different methods of constructing credible intervals will lead to the
same results \cite{Hamann:2007pi}. In the following,
we will quote the standard deviation as a measure of uncertainty in
parameter $\theta$,
\begin{equation}
\sigma_\theta = \sqrt{\frac{1}{N} \sum_{i=1}^N \left(\theta_i -
    \bar{\theta} \right)^2},
\end{equation}
where $i$ runs over the points of the Markov chain and $\bar{\theta}$
is the mean of the $\theta_i$. This quantity corresponds to the width
of the usual minimal $68\%$ credible interval.

%%%%%%%%%%%%%%%%%%%%%%%%%%%%%%%%%%%%%%%%%%%%%%%%%%%%%%%%%%%%%%%%%%%%%%
\section{Results\label{sec4}}
%%%%%%%%%%%%%%%%%%%%%%%%%%%%%%%%%%%%%%%%%%%%%%%%%%%%%%%%%%%%%%%%%%%%%%

\subsection{Vanilla model}

\begin{table}[t]
%\begin{center}
\caption{In this table we list the relative uncertainties
  $\sigma_\theta/\bar{\theta}$ of the 
  six parameters of the vanilla model, given in per cent. 
\label{table:vanillaerrors}}\vskip5mm \hskip35mm
\footnotesize{\begin{tabular}{lcccc}
 \br
&{\sc planck}&{\sc planck}$_{\rm fixed}$ &{\sc cvl}&{\sc cvl}$_{\rm fixed}$\\
 \mr
 $\Omega_{\rm b} h^2$ & 0.602 & 0.607 & 0.187  & 0.149 \\
 $\Omega_{\rm dm} h^2$ & 1.14 & 1.12& 0.587 & 0.581 \\
 $h$ & 0.859 & 0.848 & 0.387 & 0.388 \\
 $z_\mathrm{re}$  & 3.43 & 3.32 & 2.00 & 2.01 \\
 $\ln(10^{10}A_\mathrm{S})$ & 0.466 & 0.465 & 0.270 & 0.274\\ 
 $n_\mathrm{S}$ & 0.366 & 0.365 & 0.198 & 0.200\\
 \br
\end{tabular}}
%\end{center}
\end{table}

The most serious potential consequence of adding extra parameters
to an inference exercise is a shift in the parameter means, i.e., a
bias in the point estimates. However, as expected, we find no such
shift for either data set, when we compare the results from the fixed
temperature analysis with the free temperature runs: the means differ
by less than $0.1\%$.

\begin{figure}[t]
%\begin{center}
\hskip25mm
\includegraphics[height=.69\textwidth, angle=270]{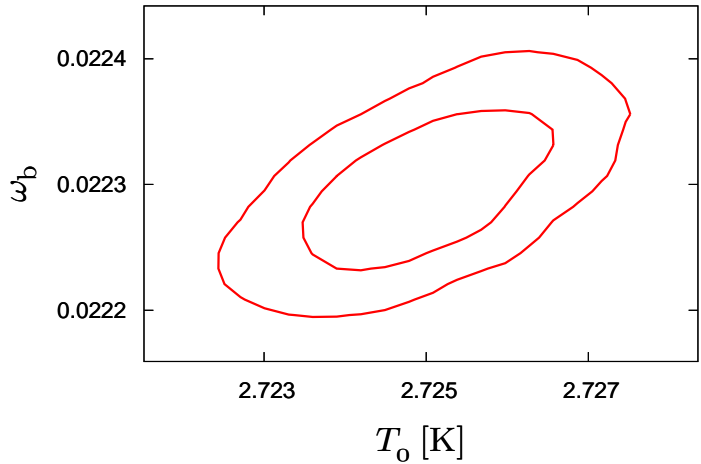}
\caption{\label{figure:tomb}
This plot illustrates the correlation between the baryon density
$\omega_{\rm b} = \Omega_{\rm b} h^2$ and the CMB temperature for a
fit of the vanilla model to the {\sc cvl} data
set. Depicted are the joint 2-dimensional $68\%$- and $95\%$-credible
contours.}
%\end{center}
\end{figure}

We do find an effect on the errors though:
the inferred uncertainties $\sigma_\theta/\bar{\theta}$ of the vanilla
model parameters are listed in Table~\ref{table:vanillaerrors}. Using
subsets of the full chains, we estimate the accuracy of these numbers
to lie at the per cent level. Apart from the baryon density, the
errors of the fixed temperature and free temperature analyses differ
by a few per cent at most, both for {\sc planck} and {\sc cvl}
data. This corresponds roughly to the expected variance of the results
for multiple runs of the same model and is consistent with a null
effect. The only significant exception is the error of the baryon
density, which for the {\sc cvl} data set is roughly $20\%$ when
taking the temperature uncertainty into account. The reason for this
increase lies in a parameter degeneracy between $T_0$ and $\omega_{\rm
  b}$, as explained in section \ref{sec2} and demonstrated in 
figure~\ref{figure:tomb}. The qualitatively
similar effect of these two parameters on the anisotropy power
spectra can also be seen in figure~\ref{fig:cls}.

This mild degradation in the sensitivity to $\omega_{\rm b}$ under 
an ideal situation indicates that the uncertainty in the CMB temperature 
measurement is, for the purpose of parameter estimation, sufficiently
well controlled.
Nonetheless, we find that a reduction in the error of the current
temperature measurement by a factor of five will bring the degradation
down to the per cent level.

It is interesting to note that had we imposed instead a temperature
prior of  $T_0 = 2.726 \pm 0.005 \ {\rm K}$ based on the original
analysis of the {\sc firas} data \cite{Mather:1993ij}, the sensitivity
of {\sc cvl} to the baryon density would degrade by as much as a
factor of 3.6 relative to the fixed temperature case. The
degeneracies between $T_0$ and the parameters $\omega_{\rm m}$ and $h$
would also manifest as a $40\%$ and a $10\%$ degradation in their
respective projected errors. Thus, through a stroke of coincidence, 
the current error in $T_0$ of $1 \ {\rm mK}$ leads to sensitivity degradations 
that are large enough to still be
detectable, and yet small enough not to significantly limit the constraining 
power of even an ideal CMB survey.

\subsection{Extended models}

While the vanilla model enjoys a large amount of popularity these
days, and is generally used as the benchmark model for parameter
estimates, it may be necessary in the future to consider extended
models with more free parameters. One such example is the primordial
Helium fraction, $Y_{\rm He}$. While current CMB data are not very
sensitive to changes in $Y_{\rm He}$, it will be necessary, already
for {\sc planck} data, to include it in the analysis
\cite{Lesgourgues:2006nd}. In fact, the projected sensitivity of the
CMB to $Y_{\rm He}$ will rival that of astrophysical measurements,
without being troubled by experimental systematics \cite{Trotta:2003xg}.

Generically, as pointed out above, adding extra parameters will tend
to increase the uncertainties on existing parameters, provided that
the data can constrain the new parameters well and barring unusual
shapes of the posterior distribution.
Our results for the vanilla model should thus be regarded as an
estimate of the maximum possible effect. As can be seen from table
\ref{table:YHeerrors}, including $Y_{\rm He}$ slightly weakens the
bounds on the baryon density due to a degeneracy with the baryon
density. As a result, the difference of the bounds of the fixed
$T_0$ and free $T_0$ analyses goes down to $\sim 10\%$. The addition
of other parameters degenerate with $\omega_b$ would further decrease
the temperature effect.

\begin{table}[t]
%\begin{center}
\caption{In this table we list the relative uncertainties
  $\sigma_\theta/\bar{\theta}$ of the 
  seven parameters of the extended vanilla+$Y_{\rm He}$ model, given
  in per cent. 
\label{table:YHeerrors}}\vskip5mm \hskip55mm
\footnotesize{\begin{tabular}{lcc}
 \br
&{\sc cvl}&{\sc cvl}$_{\rm fixed}$\\
 \mr
 $\Omega_{\rm b} h^2$ & 0.248 & 0.226 \\
 $\Omega_{\rm dm} h^2$ & 0.574 & 0.565 \\
 $h$ & 0.396 & 0.392 \\
 $z_\mathrm{re}$  & 2.05 & 1.99 \\
 $\ln(10^{10}A_\mathrm{S})$ & 0.339 & 0.338 \\ 
 $n_\mathrm{S}$ & 0.317 & 0.309 \\
 $Y_{\rm He}$ & 1.35 & 1.33 \\
 \br
\end{tabular}}
%\end{center}
\end{table}

\subsection{The dipole}

Equation~\ref{eq:diponorm} shows that there is a direct degeneracy
between the CMB dipole and the inferred value of the normalisation of
the initial power spectrum. From Table~\ref{table:vanillaerrors}, we see that the relative
error on the logarithm of the normalisation, $\ln \left[ 10^{10} A_{\rm
S}\right]$, is about 0.27\% even in the most optimistic case in which
the dipole is infinitely well known, the data is cosmic variance
limited and a minimal model is assumed.  This corresponds to a
relative error in $A_{\rm S}$ of roughly 0.9\%. Adding up this error
and the dipole error of $0.4\%$ quadratically, one expects an effect
on the error of $A_s$ of less than 10\%.  This rough estimate is
confirmed by our MCMC analysis: we find that for the minimal model and
the {\sc cvl} data set, fixing the dipole will lead one to
underestimate the error of the normalisation $A_{\rm S}$ by 8\%, while
the other parameters are affected by less than 1\%. We can thus
conclude that the {\sc firas} dipole measurement is sufficiently
accurate for the purpose of future cosmological parameter inference.

%%%%%%%%%%%%%%%%%%%%%%%%%%%%%%%%%%%%%%%%%%%%%%%%%%%%%%%%%%%%%%%%%%%%%%
\section{Conclusion\label{sec5}}
%%%%%%%%%%%%%%%%%%%%%%%%%%%%%%%%%%%%%%%%%%%%%%%%%%%%%%%%%%%%%%%%%%%%%%

We have shown that ignoring the uncertainty in the measurement of the
present CMB temperature \cite{FixMat,Mather:1998gm}
$T_0 = 2.725 \pm 0.001 \ {\rm K}$, can affect the extraction of cosmological
parameters from future data.  However, the magnitude of this effect is
rather small. While for projected {\sc planck} data it appears to be
altogether negligible, one runs the risk of underestimating the error
in the baryon density by about 20\% for an ideal, cosmic variance
limited experiment, assuming the current six-parameter vanilla
model. For the other parameters of this model, the difference is at
most at the per cent level.
An improved measurement of $T_0$, reducing the current error 
by a factor of five, would remedy this problem.
On the contrary, if one were to use the result of the original {\sc firas}
analysis \cite{Mather:1993ij}, $T_0 = 2.726 \pm 0.005 \ {\rm K}$, 
the effect of the temperature uncertainty would be much more dramatic: 
the projected error in the baryon density would increase by a factor of 3.6. 
Even the sensitivities to the matter density and the Hubble parameter 
would suffer some mild degradation. 

In the same vein we have also estimated the effect of the CMB dipole
uncertainty. We found that taking into account the dipole error
of $0.2\%$ degrades the sensitivity to the normalisation of the
primordial power spectrum by less than $10\%$ for a cosmic variance
limited experiment, compared to the case in which the dipole is
infinitely well known.

We conclude that, at least from a parameter estimation point of view,
the present precision of CMB temperature monopole and dipole
measurements is ``good enough''. However, it should be stressed that
an improved measurement of the CMB spectrum would nevertheless be a
worthwhile endeavour, for two reasons. Firstly, it offers the prospect
for detecting possible global deviations from the blackbody spectrum,
typically parameterised in terms of the Bose--Einstein and Compton
distortions $\mu$ and $y$. As pointed out in reference~\cite{FixMat},
using state-of-the-art technology, the current $95\%$ c.l.~limits of
\mbox{$|\mu|<9\times10^{-5}$} and
\mbox{$|y|<1.5\times10^{-5}$}~\cite{Fixsen:1996nj} could be improved by two
orders of magnitude. Secondly, an actual detection of the signatures
left by the process of recombination could serve as an additional,
independent probe of cosmological parameters, such as the baryon
density \cite{Chluba:2007zz}, as well as testing our understanding of
recombination physics.

%%%%%%%%%%%%%%%%%%%%%%%%%%%%%%%%%%%%%%%%%%%%%%%%%%%%%%%%%%%%%%%%%%%%%%
\section*{Acknowledgments}
%%%%%%%%%%%%%%%%%%%%%%%%%%%%%%%%%%%%%%%%%%%%%%%%%%%%%%%%%%%%%%%%%%%%%%

It is a pleasure to thank Jens Chluba, Steen Hannestad and An\v{z}e
Slosar for interesting discussions and valuable comments. We
acknowledge the use of computing resources from the Danish Center for
Scientific Computing (DCSC). JH acknowledges support by the Deutsche
Forschungsgemeinschaft under grant TR~27 ``Neutrinos and beyond''.

%%%%%%%%%%%%%%%%%%%%%%%%%%%%%%%%%%%%%%%%%%%%%%%%%%%%%%%%%%%%%%%%%%%%%%
\appendix

\section{Mock data generation and the likelihood function} \label{sec:appendix}
%%%%%%%%%%%%%%%%%%%%%%%%%%%%%%%%%%%%%%%%%%%%%%%%%%%%%%%%%%%%%%%%%%%%%%

We demonstrate in this section how to generate random realisations of 
future CMB data given some fiducial model, for the purpose of parameter 
error forecast.
The method outlined below is essentially a 
generalisation of the procedure introduced in reference~\cite{Perotto:2006rj},
and can be applied also to forecasts for, e.g.,  cosmic shear experiments.

Sky maps of the CMB are usually expanded in spherical harmonics, 
where the coefficients, or the multipole moments,
$a_{\ell m}^\mu$ in the mode $\mu$ ($\mu=T,E,\ldots$.)
receive contributions from both the signal 
$s_{\ell m}^\mu$ and the experimental noise $n_{\ell m}^\mu$,
\begin{equation}
a_{\ell m}^\mu = s_{\ell m}^\mu + n_{\ell m}^\mu.
\end{equation}
Assuming the experiment has a full sky coverage and a spatially uniform 
Gaussian noise spectrum, the total covariance matrix  
$\widetilde{C}_\ell^{\mu \nu} \equiv \langle a_{\ell m}^{\mu *}
a_{\ell m}^\nu \rangle$ is diagonal in the $\ell$ basis, and can be
written as a sum of the signal $C^{\mu \nu}_\ell\equiv \langle s_{\ell
  m}^{\mu *} s_{\ell m}^\nu \rangle$ and noise $N^{\mu \nu}_\ell\equiv
\langle n_{\ell m}^{\mu *} n_{\ell m}^\nu \rangle$ 
power spectra,
\begin{equation}
\widetilde{C}_\ell^{\mu \nu} 
= C^{\mu \nu}_\ell + N^{\mu \nu}_\ell.
\end{equation}

Given a fiducial cosmological model ${\bm \theta}_0$ and the 
noise specifications of the experiment of interest, one can calculate
$\left. C^{\mu \nu}_\ell \right|_{{\bm \theta}_0}$ and hence
$\left. \widetilde{C}^{\mu \nu}_\ell \right|_{{\bm \theta}_0}$. Random
realisations of the fiducial model can then be generated as follows:
\begin{enumerate}
\item Generate row vectors ${\bm G}_{\ell m} =\{G_{\ell m}^1,G_{\ell
    m}^2, \ldots, G_{\ell m}^n\}$,
each consisting of $n$ random numbers drawn from a Gaussian
distribution.  The number $n$ corresponds to the number of observable
modes (e.g., $\mu=T,E$ makes $n=2$).

\item The observables ${\bm A}_{\ell m}=\{a_{\ell m}^1,a_{\ell m}^2,
  \ldots, a_{\ell m}^n\}$ are defined as 
\begin{equation}
{\bm A}_{\ell m} = {\bm G}_{\ell m} {\bm L}^T,
\end{equation}
where ${\bm L}$ is a lower triangular matrix satisfying the relation
$\left.\widetilde{\bm C}_\ell\right|_{{\bm \theta}_0} = {\bm L} \cdot
{\bm L}^T$.  
The components of ${\bm L}$ can be obtained from a Cholesky
decomposition, so that the diagonal elements are given by 
\begin{equation}
L_{\mu \mu} = \left( \left.\widetilde{C}^{ \mu \mu}_\ell\right|_{{\bm
      \theta}_0} - \sum^{\mu-1}_{\rho=1} L_{\mu \rho}^2 \right)^{1/2}, 
\end{equation}
and the off-diagonal elements by
\begin{equation}
L_{\nu \mu} = \frac{1}{L_{\mu \mu}} \left(\left.\widetilde{C}^{\mu
      \nu}_\ell \right|_{{\bm \theta}_0} 
- \sum^{\mu-1}_{\rho=1} L_{\mu \rho} L_{\nu \rho} \right), \qquad
L_{\mu \nu} = 0, 
\end{equation}
with $\nu=\mu+1,\mu+2,\ldots,n$.

\item The mock power spectra are constructed by summing the bilinear
  products of $a^\mu_{\ell m}$,
\begin{equation}
\hat{C}_\ell^{\mu \nu} = \frac{1}{2 \ell + 1}   \sum^\ell_{m=-\ell}
a^{\mu*}_{\ell m} a^\nu_{\ell m}. 
\end{equation}

\end{enumerate}

\bigskip

To extract parameter errors from  the mock data we approximate the
total likelihood function ${\cal L}$ as a multivariate Gaussian in the
mock multipole moments $a_{\ell m}^\mu$. Equivalently,
\begin{equation}
\chi^2_{\rm eff} \equiv -2 \ln {\cal L} = \sum_\ell (2 \ell +1) 
\left[{\rm Tr}(\widetilde{\bm C}_\ell^{-1} \hat{\bm C}_\ell) + 
\ln \frac{|\widetilde{\bm C}_\ell|}{|\hat{\bm C}_\ell|} - n \right],
\end{equation}
where we have made use of the fact that both the mock data 
and the noise power spectra are  diagonal in the $\ell$ basis. 
We approximate the effect of  the mandatory sky cut near the galactic plane
with a fudge factor $f_{\rm sky}$,
\begin{equation}
\label{eq:chi2}
\chi^2_{\rm eff} = \sum_\ell (2 \ell +1) \ f_{\rm sky} 
\left[{\rm Tr}(\widetilde{\bm C}_\ell^{-1} \hat{\bm C}_\ell) + 
\ln \frac{|\widetilde{\bm C}_\ell|}{|\hat{\bm C}_\ell|} - n \right],
\end{equation}
where $f_{\rm sky}$ stands for the actual fraction of the sky observed after
the cut.

Finally, we note that it is also possible to perform a forecast using 
the fiducial $\left.\widetilde{C}_\ell^{\mu \nu} \right|_{{\bm \theta}_0}$
instead of a random realisation of the fiducial model, i.e., one can set
$\hat{\bm C}_\ell$ equal to $\left.\widetilde{\bm C}_\ell
\right|_{{\bm \theta}_0}$ 
in equation~(\ref{eq:chi2}).
This amounts to considering an average over an infinite number 
of independent realisations of the same fiducial model, and produces  
essentially similar error estimates as the more complicated procedure
outlined above.

%%%%%%%%%%%%%%%%%%%%%%%%%%%%%%%%%%%%%%%%%%%%%%%%%%%%%%%%%%%%%%%%%%%%%%
\section*{References}
%%%%%%%%%%%%%%%%%%%%%%%%%%%%%%%%%%%%%%%%%%%%%%%%%%%%%%%%%%%%%%%%%%%%%%

\end{document}